\documentclass[amssymb,preprint,preprintnumbers,nofootinbib,superscriptaddress]{revtex4}

\usepackage{graphicx}
\usepackage{latexsym}
\usepackage{amsfonts}
\usepackage{epstopdf}

\begin{document}

\title{Revealing shifts from mastery of knowledge to problem solving in assessments of a tertiary physics programme}

\author{Alan~S.~Cornell}
\email[Email: ]{acornell@uj.ac.za}
\affiliation{Department of Physics, University of Johannesburg, PO Box 524, Auckland Park 2006, South Africa.}

\author{Kershree~Padayachee}
\email[Email: ]{kershree.padayachee@wits.ac.za}
\affiliation{Science Teaching and Learning Unit, Faculty of Science, University of the Witwatersrand, Wits 2050, South Africa.}

\begin{abstract}
There is an increasing pressure for lecturers to work with two goals. First, they need to ensure their undergraduate students have a good grasp of the knowledge and skills of the intellectual field. In addition, they need to prepare graduates and postgraduates for careers both within and outside of academia. The problem addressed by this paper is how assessments may reveal a shift of focus from a mastery of knowledge to a work-focused orientation. This shift is examined through a case study of physics and the sub-discipline of theoretical physics as intellectual fields. The evidence is assessment tasks given to students at different points of their studies from first year to doctoral levels. By examining and analysing the assessment tasks using concepts from Legitimation Code Theory (LCT), we demonstrate how the shifts in the assessments incrementally lead students from a pure disciplinary focus to one that enables students to potentially pursue employment both within and outside of academia. In doing so, we also highlight the usefulness of LCT as a framework for evaluating the preparation of science students for diverse workplaces.     
\end{abstract}

\maketitle

%
%
\section{Introduction}

\par The higher education landscape has shifted dramatically in the past 20 years. Worldwide, staff and student demographics have changed and the very purpose of higher education is being challenged. The world of work is also currently in a state of flux, with requirements for new ``21st century'' skills and competency emerging constantly (Bao and Koenig, 2019). The result is increasing pressure for lecturers to shift curricula that focus tightly on disciplinary knowledge and procedures and provide greater opportunities for students to develop broader, cross field competencies that will equip students for a wide range of potential careers (Ashwin and Case, 2018; Roberts, 2015).

\par The branching out of the academic pipeline, with graduates (especially those with higher degrees) entering workplaces that are vastly different from the academic environment in which they have been prepared. This indicates the need for graduates to develop both in-depth and highly sophisticated understandings of the knowledge of their disciplines, in order that they be able to apply this knowledge to a broad range of contexts and fields of practice (Acker and Haque, 2017). It may also have implications for the type of identity graduate students develop, based on their intrinsic motivations and their intended career paths. Higher education therefore, needs to adapt in response to the shifting socio-economic pressures and requirements to enable the development of graduates with flexible dispositions that would allow for easier integration into new contexts and easier application of disciplinary knowledge in fields of practice beyond the academe.  However, while undoubtedly necessary, it is possible that such shifts may present challenges for the development and training of students in pure, hard science disciplines such physics and specialised sub-disciplines such as theoretical physics. In this paper, we firstly analyse what is valued in the discipline at different stages of theoretical physics education. We then draw on the analysis to highlight the epistemological shifts that occur in the process of theoretical physicist identity development as students move from first year to PhD students in the discipline. Lastly, we discuss how these shifts incrementally develop students? abilities to create new knowledge and apply the knowledge of physics and theoretical physics in other fields over time.  

%
%
\section{Progression in the learning of physics}

\par Physics is regarded as a discipline where progress is attained by seeking greater levels of abstraction, such as universally applicable laws that govern natural phenomena. Terms are generally precisely defined and there are similar bodies of knowledge taught in introductory physics courses in different institutions. On these grounds, physics can be described as having a generally hierarchical knowledge structure where there is broad consensus on precisely defined concepts and relationships between them.  In fields like these, knowledge-building happens through the integration of foundational concepts to more abstract and universal claims (Bernstein, 1999; 2000). As such, the undergraduate physics curricula tend to follow similar topics and sequences across different settings. This curriculum format is deemed necessary to ensure that students become familiar with the foundational knowledge and procedures of the discipline. 

\par First year physics classes generally include students registered for engineering and medical degrees, as well as students intending to major in physics or chemistry. An introductory course in physics would usually include topics relating to Newtonian mechanics, waves and optics, electromagnetism and so on. At this level, the curriculum would also include the development of thinking and reasoning skills involved in enquiry, experimentation, and evidence evaluation. The transition to the more specialised sub-disciplines of theoretical physics occurs in the latter years of the undergraduate programme, a transition requiring greater conceptual understanding and knowledge integration. Compared with general physics courses, specialised sub-disciplines such as theoretical physics focus on the more abstract generation of theoretical models, require more complex mathematical knowledge, and demand a higher level of logic and critical thinking. In this respect, critical thinking skills are expanded to include inference and argumentation, the ability to systematically explore a problem, formulation and testing of hypotheses, manipulation and isolation of variables, and observation and evaluation of consequences (Zimmerman, 2000). In the process of becoming increasingly specialised, experiential aspects of physics are left aside in favour of a greater emphasis on conceptual understanding and the development of particular mental models and procedural knowledge.  

\par There is, accordingly, a gradual development of increasingly complex conceptual understanding, and that occurs in the education of theoretical physics students, evolving from conceptual knowledge acquisition to understanding and comprehension of the knowledge and the specific procedures that are underpinned by these. That is, students are gradually inducted into the discipline and the practice of knowledge production, gaining foundational knowledge and `learning to think like a physicist' during undergraduate studies (Van Heuvelen, 1991; Conana, 2016; Conana, Marshall and Case, 2020). Lecturers thus initially aim to construct disciplinary knowledge in a hierarchical manner, by beginning with hands on experiential work through laboratories and practicals, where this quickly becomes a repeated testing of invented threshold concepts (Wisker, 2018) such as Newton's laws etc., to see if these always hold-up under scrutiny, as well as the constant development of students' understanding of complex representations and their uses in physics. 

\par This repeated testing and questioning that characterises undergraduate studies is a key aspect of the physicist's identity, founded on Dewey's notion of `competent inquiry' (Dewey, 1938, loc. cit. Towne and Shavelson, 2002). It can also be associated with Popper's ``critical rationalism'', amongst other scientific philosophies (Towne and Shavelson, 2002). The key point is that the nature of the scientific laws that are sought to be formulated are abstract ones that must constantly be tested. In the context of theoretical physics, this is captured in the lecture series entitled ``Character of Physical Law'', by Feynman (1965), in how he saw ``physical laws''. That is, he suggested a theoretical physicist attempts to write down, in the language of mathematics, a simple and beautiful explanation for nature. Feynman points out, however, that this is, at best, a model, one that will only have a certain regime of applicability. Physicists test the range of this model, whilst refining and improving, in a search for new laws and broader or deeper understandings and applications, making this the cornerstone of physics education. 

\par Physics education, therefore, needs to be underpinned by an appreciation for the nature of science and a solid conceptual foundation from which the skills of abstraction, critical thinking and new knowledge production can emerge. The undergraduate physics curriculum, therefore, requires a solid conceptual foundation that includes critical thinking skills for subsequent abstraction and knowledge production. The role of the physics lecturer is consequently to facilitate access to, and engagement with the concepts in a way that leads students to ever more complex levels of abstraction and criticality, and in the process, gradually shaping a scientific identity and a ``trained gaze'' (Maton, 2014, p. 186) of a theoretical physicist. A gaze is a specialised way of seeing, perceiving and thinking in the knowledge practices of Physics and has been developed through interactions, including assessment tasks that emphasise the mastery of specialist knowledge and skills. 

\par Conana (2016) and Conana et al. (2020) explain how a process of knowledge-building of specialised knowledge and skills typically occurs in first year physics lectures, and how a deliberate waving between everyday knowledge enhances understanding of physics concepts and procedures. This approach enables students to see the application and broader relevance of concepts from the discipline, making engagement with the concepts more meaningful for students studying physics as a pre-requisite for further study in the related fields of Medicine and Engineering. It also triggers interest and curiosity in students who intend majoring in physics. It can be argued, however, that it is summative assessments that ultimately drive students' focus in the curriculum and determine students' success, by signalling to students what knowledge mastery and graduate attributes are valued at each stage of study (Boud \& Falchikov, 2007). 

\par In the case of physics and theoretical physics, the shifts towards increasing complex conceptual understanding, greater levels of intellectual independence and demands for disciplinary ways of thinking and reasoning as key outcomes of the courses, should, therefore, be reflected in the demands of assessment tasks. It is, therefore, critical in our endeavour, to analyse how the physics and theoretical physics assessment demands vary to promote the students' ability to develop and then apply the knowledge of these disciplines in different disciplines and work contexts. Analysing knowledge practices, like shifts in assessment tasks in an intellectual field like physics, however, requires a conceptual framework that can reveal changes in what is valued for achievement over different levels. It is for this reason we turn now to the conceptual tools offered by the Specialization dimension of Legitimation Code Theory (LCT) (Maton, 2014). 

%
%
\section{Epistemic relations and learning Physics in Higher Education}

\par LCT offers a sociological approach to researching and analysing knowledge practices. It offers a multidimensional conceptual toolkit which enables ``both the exploration of knowledge-building and the cumulative building of knowledge'' allowing ``knowledge practices to be seen, their organizing principles to be conceptualized, and their effects to be explored'' (Maton, 2014, p. 4). This is achieved through the selective application of the different dimensions of LCT, each of which comprise ``a series of concepts centred on capturing a set of organizing principles underlying dispositions, practices and contexts'' (Maton, 2016, p. 11). The Specialization dimension provides concepts that reveal the extent to which the basis of achievement in a practice or intellectual field lies in the mastery of a body of specialised knowledge and/or the acquisition of specialised dispositions, ways of thinking or being. Conceptual tools from LCT have been used to analyse assessment tasks in chemistry (e.g., Blackie, 2014), student understandings in physics (e.g., Georgiou, 2016) and knowledge-building in earth sciences (e.g., Maton \& Doran, 2021). In this paper we use the strength of epistemic relations and knowledge-based insights to consider how the focus on knowledge and procedures shifts in the assessment of physics at various levels in a higher education setting.

\par The Specialization dimension of LCT arises from the principle that knowledge practices are orientated towards a part of the world and are enacted by actors. This principle sets up two relations, an epistemic relations (ER) between the practice and its objects of study, and social relations (SR) between the practice and its subjects. epistemic relations are defined by relations of the knowledge to its object of study (i.e., the relation between the knowledge and the part of the world/disciplinary field in which the body of knowledge is directed). In contrast, social relations are defined by relations between the knowledge practices and the subject (i.e., the relations between those making the claim to the knowledge). 

\par When the epistemic and/or social relations are highly significant as a basis for achievement in the knowledge practice, relations are regarded as being stronger along a continuum of strengths. Similarly, when the relations matter less as a basis for achievement, the relations are said to be relatively weaker. Different intellectual fields and different fields of professional practice may thus be characterised by the relative strengths or weaknesses placed on epistemic and social relations which may be indicated by ``+'' or ``-'', respectively.

\par The concepts and codes of Specialization provided the language and tools to analyse the logics that underpin the development of theoretical physicists, as a starting point for a deeper exploration of what is valued as the basis of achievement in this field. Moreover, in this study we focus on how cognitive demands in assessment tasks shift over time as students move from undergraduate to post graduate physics courses. We are especially interested in what knowledge insights are developed through shifts in the epistemic relations of the knowledge practices of physics. For this reason, we focus on the epistemic plane of LCT which gives the means to analyse how epistemic relations change strength and form in the assessments at different stages of a theoretical physicists' education, both at undergraduate and postgraduate level. 

\par The epistemic relations of knowledge practices may be specialised according to the extent to which the field determines legitimate objects of study, and the procedures by which these objects are investigated (Maton, 2014). The epistemic plane enables deeper consideration of the object of study of one field relative to another, through considerations of the relative strengths of the ontic and discursive relations. Ontic relations (OR) describe the extent to which the object of study in the field is bound to the field, i.e., the ``what'' of its knowledge practices. The ontic relations, therefore, reflect the strengths of relations between knowledge practices within the intellectual field and their demarcated object/s of study (Maton, 2014). Discursive relations (DR), however, refers to the extent to which there are clear and uncontested procedures that determine how those object/s are studied. Discursive relations therefore, set up relative strengths between knowledge practices associated with a particular intellectual field (in this study, physics) and procedures used to investigate problems in other knowledge practices (Maton, 2014). In short, discursive relations consider the ``how'' of knowledge practices.

\begin{figure}[h]
\includegraphics[width=0.4\textwidth]{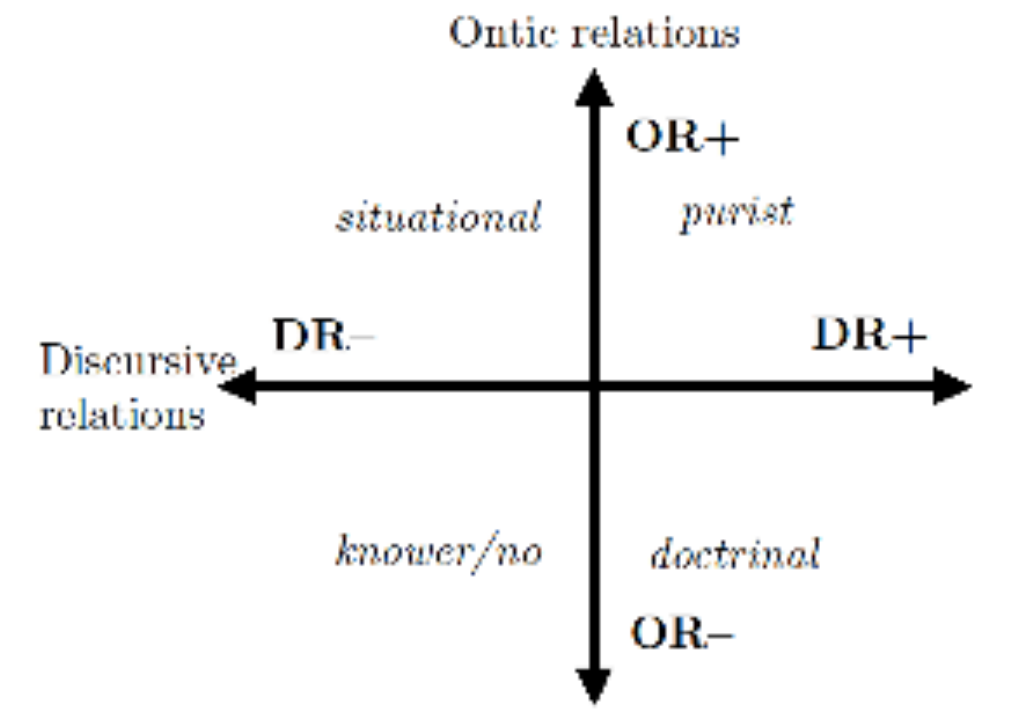}
\caption{\it The epistemic plane of LCT, showing four codes that form the basis of knowledge-based insights.}\label{Fig:1}
\end{figure}

Maton (2014) explains that a field's ontic and discursive relations can also be combined to illustrate the continuum of strengths of relations on the epistemic plane, resulting in four analytically distinct quadrants representing four insights of epistemic codes, viz., purist, doctrinal, knower/no and situational (Figure \ref{Fig:1}). 
\begin{itemize}
\item Knowledge practices dominated by purist insights are generated by fields that ``strongly bound and control both legitimate objects of study and the legitimate approaches'' used to study them (Maton, 2014 p. 176). Such fields where the object of study and approaches are strongly controlled by the field, are said to be characterised by strong ontic fidelity and procedural purism. 
\item In comparison, knowledge practices characterised by doctrinal insights (OR-, DR+) are generated by fields where the object of study is less determined by the field (weaker ontic fidelity) but specialised procedures are of greater significance for achievement in the field (stronger discursive relations). 
\item Situational insights (OR+, DR-), on the other hand, develop in knowledge practices where the object of study is controlled by the field but there is greater potential for application of multiple approaches or procedures to address a particular issue or problem. 
\item Knower insights or no knowledge-based insights, in contrast, are generated in practices where neither the objects of study nor the procedures used to study them are dictated by the field. These insights are characterised by generally weaker ontic and discursive relations.   
\end{itemize}

\noindent The epistemic plane thus provided a useful analytical framework for analysing the assessments in order to reveal the knowledge practices that are foregrounded in the education of theoretical physicists, and the extent to which the programme prepares students for diverse workplaces. 

%
%
\section{Methodology}

\par This paper offers a qualitative analysis of assessment tasks designed for undergraduate Physics courses as well as assessments from theoretical physics Honours courses and Masters and PhD projects in theoretical physics. These were designed by Prof A, a theoretical physicist specialising in high energy particle physics and gravitational physics at an urban university in South Africa. Supporting data was also derived from an interview between Prof A and Author 2, an educational developer based in the Faculty of Science of another urban university in South Africa, in which Prof A reflected on the nature of the assessments and the reasons for the assessment strategies adopted at different levels of study. The assessment artefacts and the transcript of Prof A's reflection on the types of questions were analysed using a soft eyes approach to determine the dominant Specialization code. This was followed by a more rigorous confirmation of the dominance of the Specialization code using a translation devise that was developed to identify and distinguish between epistemic and social relations and the relative strengths of these. 

\par Having determined that epistemic relations were emphasised much more strongly than social relations in our data, a second translation device (Table \ref{Tab:1}) was created and applied to evaluate the relative strengths of the ontic and discursive relations.

\begin{table}[h]
\caption{\it The translation device used to evaluate the relative strengths of the ontic and discursive relations in summative assessments and in the transcript of the lecturer interview}\label{Tab:1}
\centering
{\bf Ontic Relations (OR)}
\begin{tabular}{|l|l|l|l|}
\hline
{\bf Theoretical} & {\bf Code} & {\bf Indicators} & {\bf Example} \\
{\bf concept} &&& \\
\hline
Knowledge & OR++ & What is being studied & how matter moves and  \\
& (very strong ontic & (disciplinary knowledge). & interacts in an idealised \\
& relations) & Notion of right knowledge. & way \\
&& Objects of study are tightly & \\
&& defined, strongly bounded & \\
&& very strongly classified. &\\
\hline
& OR+ & Emphasis on understanding of & how matter moves and \\
& (stronger ontic & physics-based concepts and & interacts in everyday \\
& relations) & principles in settings beyond & settings \\
&& physics. &\\
\hline
& OR- & Less emphasis on disciplinary & Example:  the interactions \\
& (weaker ontic & knowledge with concepts and & and relationships of objects \\
& relations) & contexts beyond physics. & beyond the focus of physics \\
\hline
\end{tabular}
{\bf Discursive Relations (DR)}
\begin{tabular}{|l|l|l|l|}
\hline
Procedures & DR++ & Disciplinary procedures are the & Scientific method, testing \\
& (very strong & basis of legitimacy. Few & hypothesis through empirical \\
& discursive & connections and/or integration & observation and controlled \\
& relations) & of other knowledges. & experimentation \\
\hline
& DR+ & Integration/application of and & \\
& (stronger & procedures from other subdisciplines &\\
& discursive & within the same field or &\\
& relations) & closely related scientific fields. &\\
\hline
& DR- & Possible application of a & Methods that lie beyond \\
& (weaker & wide range of procedures & experimental methods, \\
& discursive & from various disciplines & analysis of observed or \\
& relations) & (beyond closely related fields) & measured variables. \\
\hline
\end{tabular}
\end{table}

\noindent A limitation of the study presented here is that we focus on the analysis of questions from one summative assessment in each of the undergraduate and Honours years of study. The choice of assessment question presented in the findings that follow were chosen by the lecturer who set them as being exemplars of the kinds of questions typically posed to students at each level.  

%
%
\section{Findings}

\par The first phase of the analysis using both the soft eyes approach and the application of the translation device to distinguish the types of questions and the relative strengths of the social and epistemic relations in the assessments, confirmed the existence of stronger ER and the dominance of the knowledge code in all the assessments at all levels. No questions required explicit demonstration of particular knower attributes, the focus lying instead on the demonstration of conceptual and procedural understanding.  This was expected and is supported by the literature on physics curricula (Bernstein, 1999) and studies on the teaching of undergraduate Physics (Conana et al., 2020). The interview with Prof A also emphasised concepts and procedures assessed at different levels and the reasons for the incremental shifts, with little mention of knower dispositions. Social relations in the data were consequently much weaker overall, compared with the strength of the epistemic relations.  

\par
\begin{figure}[h]
\includegraphics[width=0.6\textwidth]{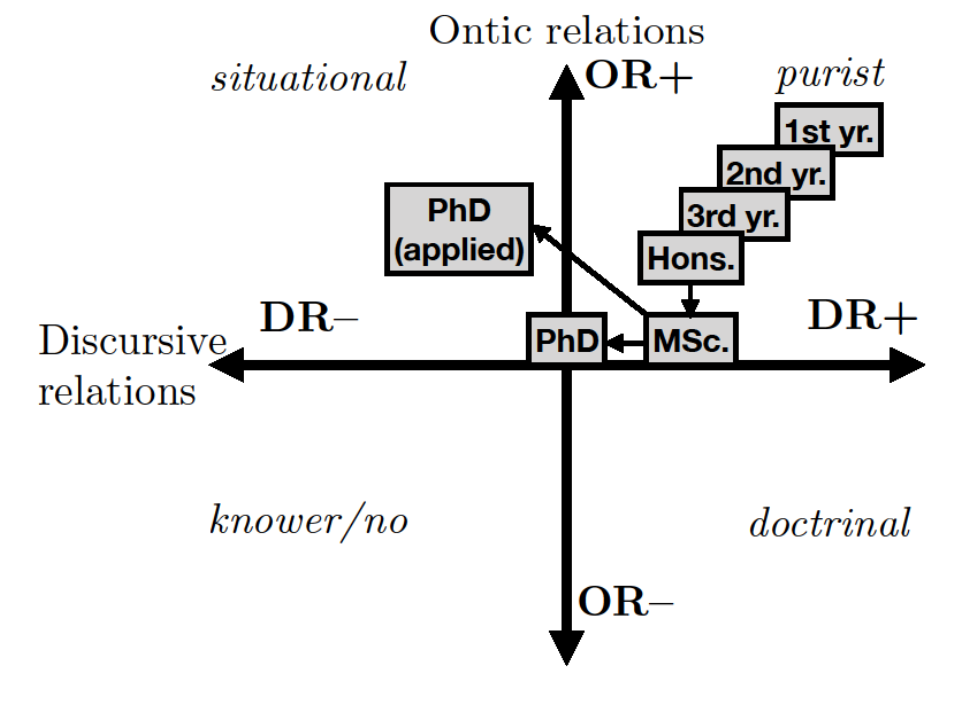}
\caption{\it A visual representation of the shifts in the strengths of ontic and social relations in the epistemic plane (Maton (2014)), evidenced in final summative assessments and postgraduate projects, as well as the lecturer's reflections on what was foregrounded in these assessments.}\label{Fig:2}
\end{figure}

 The confirmation of stronger epistemic relations and weaker social relations in all the assessment tasks enabled us to assign a `knowledge code' to the tasks in the initial analysis of the assessments. This raised the question about what knowledge-based insights were developed through the assessment tasks. This allowed for progression to Phase 2, which involved the examination of the topography of the knowledge practices found in the assessments for different years of study, on the epistemic plane. This examination revealed the dominance of purist insights (OR+, DR+) at all levels except PhD, where students may either choose to remain within the field of theoretical physics or opt for projects requiring the application of disciplinary knowledge into different intellectual fields, the latter resulting in a shift into a situational insight (Figure \ref{Fig:2}).

\noindent The shifts illustrated in Figure 2 can be best exemplified in the following excerpts from the data, where, as highlighted in Table 1, at first year level problems are highly idealised and focused on simple one idea problems. Example 1 illustrated this type of problem, in the form of a multiple-choice question: \\

{\bf Example 1:}\\
{\it 3685 g + 66.8 kg = \\
A) 7.05 x 109 mg	B) 7.05 x 104 mg 	C) 7.05 x 107 mg 	D) 7.05 x 106 mg 	E) 7.05 x 105 mg}\\

\noindent In this example and all others that follow, the basis of achievement lies entirely in the mastery of units of measurement (stronger epistemic relation) and does not require any specialised disposition or attribute of the students to succeed (very weak social relation). The emphasis is on the recognition of the differences in the units of mass and student's ability to convert to the same unit and only then add the values. The focus of knowledge is clearly demarcated and the four options of possible answers shows that students must convert both masses to milligrams. In addition, students are required to know how to use the mathematics notion of powers of ten. The knowledge focus of the question thus relies on both a mastery of knowledge from physics and knowledge from mathematics (as mentioned by Feynman (1965)), and reiterated by Prof A in the interview:

\noindent {\it ``Mathematics is recognised as the language of Physics $\ldots$ the sub-discipline of theoretical physics is always focussed on a mathematical model building and testing of Physics.''}

\noindent The example above also highlights that the procedures required to answer this question are specified and strongly bounded and so the discursive relations are strong (DR++). Other questions in the first-year assessment papers similarly focused on one specific concept or procedure at a time, with very specific ``right answer'' expected, as shown in example 2. \\

{\bf Example 2:}\\
{\it In an inelastic collision between two objects: \\
A) the kinetic energy of the system is conserved.\\
B) the momentum of each object is conserved. \\
C) both the momentum and the kinetic energy of the system are conserved. \\
D) the momentum of the system is conserved but the kinetic energy of the system\\ \hspace{0.5cm} is not conserved. \\
E) The momentum of the system is not conserved. }\\

\noindent Questions like example 2 exemplify the statement by Prof A that at first year level, ``Memorisation is probably enough to pass''. The first year assessment questions thus, tend to be characterised by very strong ontic relations (i.e., high ontic fidelity) as well very strong discursive relations (i.e., procedural purism), clustering tightly on the top right-hand corner of the purist quadrant of Figure \ref{Fig:2}. 

\par As students progress through subsequent years of study, the data shows that the assessment questions become more realistic and applied in wider and wider contexts, where discussion and conceptual understanding is tested more. For example, at the second year of study, a student may be asked the question shown in part b of Example 3. \\

{\bf Example 3:}\\
{\it (a) Show that the wave function
$$\psi (x) = A e^{m \omega x^2/2\hbar}$$
solves the time-independent Schrodinger equation for a harmonic oscillator potential 
$$V(x) = \frac{1}{2}m\omega^2 x^2$$
and find the corresponding energy.\\
(b) The allowed energies of the harmonic oscillator are
$$E_n = \left( n + \frac{1}{2} \right) \hbar \omega , \;\; n = 0, 1, 2, \ldots$$
Explain why the energy you found in the first part of the question (a), isn't among the allowed energies. }\\

\noindent Example 3 would require explaining why, among the allowed energies, a particular result is not physically acceptable. Questions of this nature require demonstrating an understanding of the application of the theory to a physical, though still quite abstract, situation. 

\par Progressing further along the qualification, students choosing to major in Physics are likely to encounter fewer questions based on single concepts and procedures, with questions shifting towards a wider application of concepts and procedures and increasingly, problems requiring problem solving, formulating inferences and making deductions, as shown in Example 4 below. 

\begin{figure}[h]
{\bf Example 4:}\\
\includegraphics[width=0.9\textwidth]{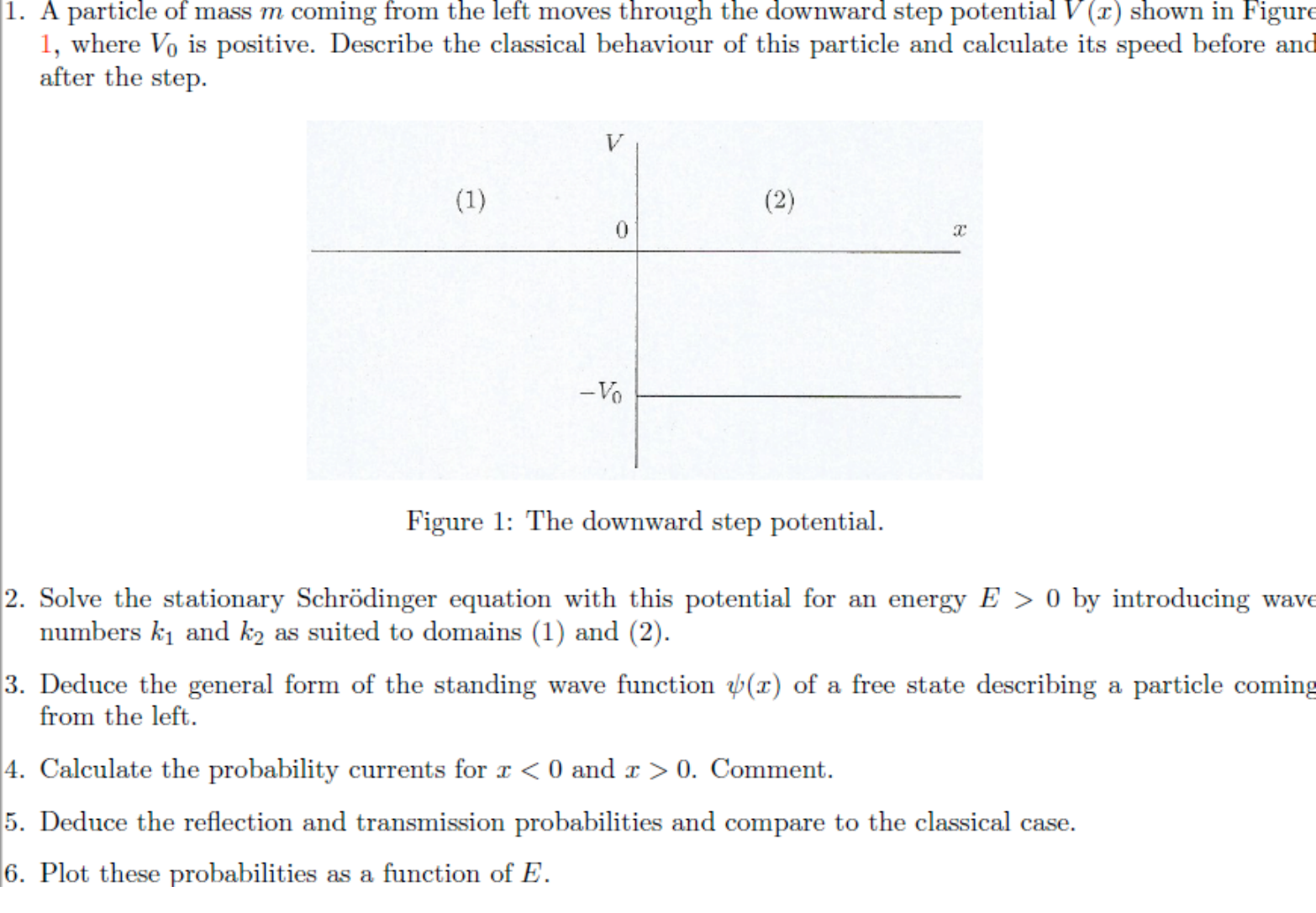}
\end{figure}

\noindent Here students are required to compare quantum and classical situations, and then to discuss and deduce behaviour from this. This is particularly evident in part 5 (i.e., `Deduce the reflection and transmission probabilities and compare to the classical ones'). The ontic and discursive relations are, therefore, both slightly weakened compared to first year questions, but overall, the purist code still dominates at this level of study. 

\par At the Honours level, there is a further weakening of the ontic and discursive relations in the summative assessment questions. Example 5, for instance, requires students to apply the techniques and tools of particle physics to a condensed matter system, a topic lying outside the usual scope of such a course. As explained by Prof A, with an example like this ``students would not have seen principles of particle physics being applied to a condensed matter system''.  A question of this sort presents an opportunity for students to apply the conceptual knowledge gained in the course in a sight-unseen way, to a different field, and ``any notions of compartmentalised knowledge are going out the window. Students need to draw from many previous concepts without explicit prompting.''

\begin{figure}[h]
{\bf Example 5:}\\
\includegraphics[width=0.9\textwidth]{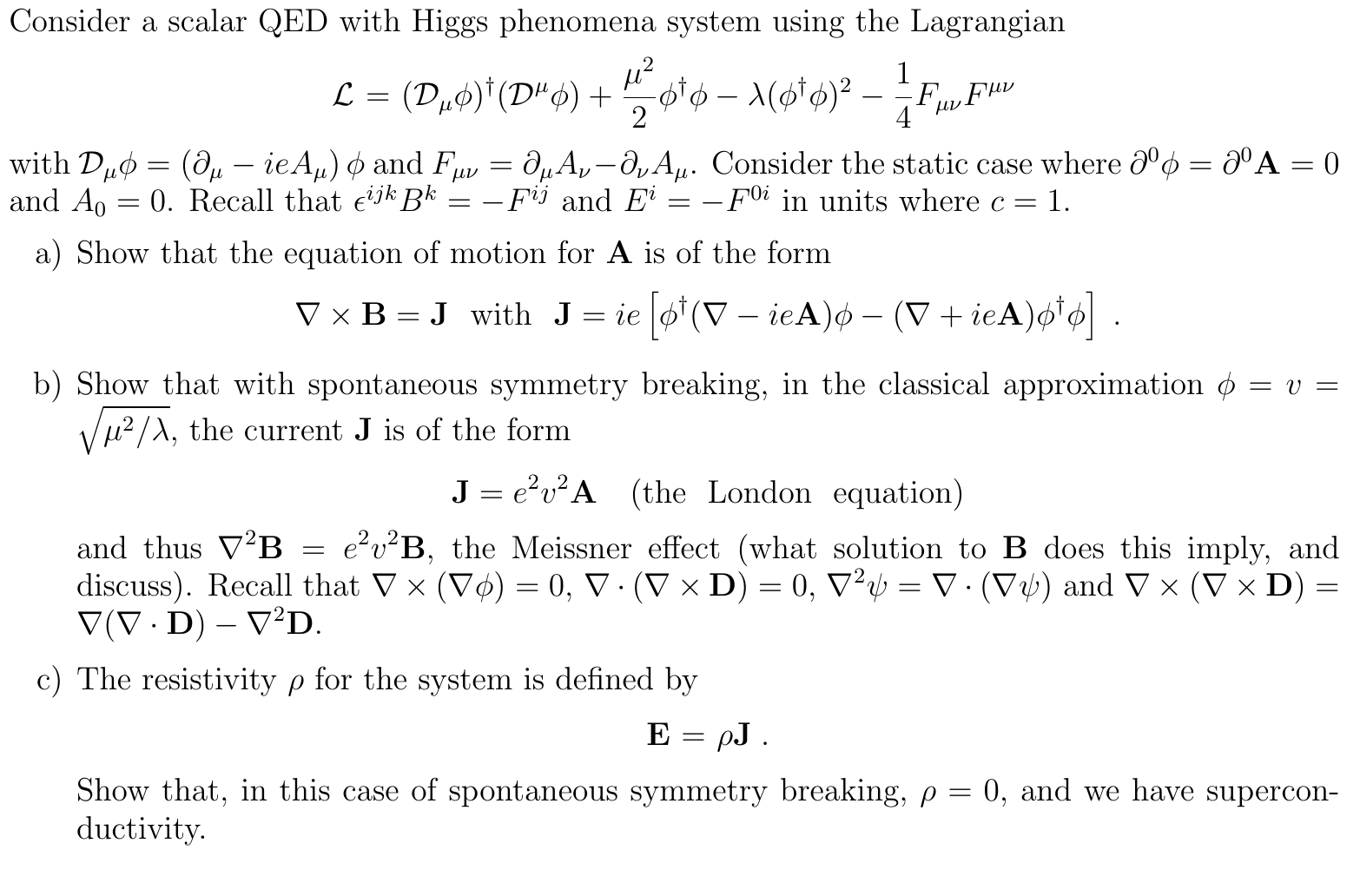}
\end{figure}

\par Example 5 illustrates that in Honours, the problems used in assessments require students to draw on multiple concepts and procedures from one field and to potentially apply these in a neighbouring or related field. The ontic and discursive relations are thus, substantially weaker than in undergraduate assessments.

\par The progressive weakening of ontic and discursive relations is further evidenced in the types of Master's and PhD projects supervised by Prof A. A Master's level dissertation, for example, has the student ``studying the quasinormal modes of black holes using artificial neural networks''. Such a study requires the integration of concepts and procedures from computer science and data analysis with advanced ideas from gravitational physics, demonstrating the weakening of ontic and discursive relations. Nevertheless, the guiding principles are based in physics and mathematics, hence the overall insight remains purist. 

\par At PhD level, the progressive widening of scope of application would be advanced further to situations where problems are now completely real world, using concepts and procedures taught in physics but applied elsewhere or in multidisciplinary ways. This is exemplified in the comment by Prof A quoting that ``a good physicist should be able to solve a problem 6 different ways''. The expectation is that the PhD student should be able to respond in seminars to questions posed by experts from different disciplines by drawing on a broad range of concepts and procedures gained in previous years of undergraduate and graduate study.  Such expectations of PhD students to creatively and innovatively apply knowledge to a range of problems in different fields is seen in the example of a PhD graduate of Prof A who completed a PhD on black hole physics and subsequently used the machine learning techniques learned in the PhD to produce, in collaboration with medical specialists, a cardiology study, a cross-disciplinary research article accepted for publication, and subsequently employment in industry. This PhD example highlights two important points: 1) the weakening of discursive relations towards procedural relativism, to the extent that there is now a shift towards a situational insight; 2) the value placed on problem solving and critical and creative thought as core outcomes for a theoretical physicist graduate at PhD level. 

%
%
\section{Discussion}

\par The examples from assessments cited above, together with Figure \ref{Fig:2}, demonstrate that the knowledge-building process in physics begins with a strongly classified first year curriculum with high ontic fidelity and procedural purism to ensure that students gain the requisite conceptual foundation needed for subsequent abstraction and knowledge production. As students demonstrate a certain degree of mastery of these foundational concepts, they are able to progress into the more specialised sub-disciplines of physics such as theoretical physics, where a different type of mastery, viz., application and conceptual integration, is valued as the basis of achievement (Figure 2). Further study develops the students' abilities to integrate and apply concepts in increasingly complex ways, in ever wider contexts, gradually developing the skills of ``competent enquiry'' described by Dewey and required of all scientists. There is hence, a gradual process unfolding over the different levels of study in which students are exposed to both the depth and breadth of the discipline over time. It is also in keeping with the notion of inducting students into the discipline through internalisation of foundational knowledge and gradually `learning to think like a physicist' (Van Heuvelen, 1991; Quan, 2017). 

\par This study also demonstrates, through the exploration of the relative strengths of ontic and discursive relations foregrounded in the assessments, how the assessments are constructively aligned with the intended outcomes of the curriculum (Biggs and Tang, 2015). The findings show that the curriculum structure underpinning the learning journey to becoming a professional theoretical physicist is one that is grounded in mastery of the concepts and procedures of physic and the closely related intellectual field of mathematics. We suggest that it would be worthwhile for this incremental approach to knowledge-building (and the intrinsic connection with mathematics) to be explained to students, so that they are able to ``see'' the learning journey ahead and engage accordingly.

\par The findings also highlighted the embedded expectation that physics graduates should, in the course of their undergraduate and graduate studies, develop the ability and skills to think critically (i.e., they should concurrently develop the skills to systematically explore a problem, formulate and test hypotheses, manipulate and isolate variables, and observe and evaluate consequences (Bao and Koenig, 2019). It is these skills that would enable graduates to transition into fields of practice outside of academia (Wang, 2018). Interestingly, the assessments and the interview data suggest a view of critical thinking as a set of skill that develops as a result of engagement with disciplinary concepts and procedures in increasingly complex ways (aligning with the views of Bao and Koening (2019) and Zimmerman (2000)), rather than the possibility of critical thinking as an innate ability or knower attribute. We note this as a potential avenue for further exploration in the next phase of the study, which will seek to both deepen our understanding of assessment in relation to knowledge building in physics and related science disciplines.

\par The analysis of the assessments in this case study using the epistemic plane and knowledge insights of LCT also enabled us to confirm that critical thinking and cross-field application of disciplinary knowledge is indeed incrementally developed in physics and theoretical physics courses and programmes. However, the analysis also revealed that application and cross field integration of disciplinary knowledge is only explicitly emphasised and addressed from the second year of study, and only foregrounded in the higher degrees. The analysis, therefore, also revealed aspects of the assessments that could potentially be improved by greater clarity and clearer communication of intended outcomes. Making the underpinning ``rules of the game'' more visible and explicit to students at every level of study could lead to enhanced learning and improved student performance, especially for `cue sensitive' student who, as pointed out by Gibbs (2010), may strategically direct their study efforts based on perceived signals of value in assessment.  

%
%
\section{Conclusion}

\par This study, conducted using LCT, enabled us to examine the underpinning logics of the assessments in Physics and Theoretical Physics, thereby allowing us to reveal the extent to which the undergraduate and postgraduate curricula are aligned for the preparation of Theoretical Physics students for diverse workplaces. The examination of ontic and discursive relations enabled us to identify and reveal the underpinning logics of the summative assessments, and to thereby demonstrate how the assessments shift students from a pure disciplinary focus to one that elicits potential for employment both within and outside of academia. The findings also showed that while students exiting with the apex qualification of the PhD should be able to apply the disciplinary knowledge to solve a wide range of problems in unrelated fields, students exiting with the base qualification of the Bachelors degree, may not be as well equipped to apply their knowledge in unrelated contexts. This has important implications for students, who should be advised during the course of the undergraduate qualification of the potential limitations of exiting at the level of the Bachelors degree. Even so, this study demonstrates that graduates who do exit at this level will have developed the capacity for critical thinking skills. 

\par Lastly, we note that while limited in scope and the extent of generalisability, this case study highlights the potential value of the concepts and tools of LCT to reveal the extent to which assessments are constructively aligned to achieve mastery of knowledge and responsiveness of curricula to the needs of employers and society. 

%
%

\section*{Acknowledgements}

\noindent ASC is supported in part by the National Research Foundation of South Africa (NRF). We are grateful to Lee Rusznyak (Wits LCT Hub, University of the Witwatersrand, Johannesburg) for her intellectual input and guidance in developing this paper.

%
%
\section*{References}

\noindent Acker, S., and Haque, E. (2017). Left out in the academic field: Doctoral graduates deal with a decade of disappearing jobs. {\it Canadian Journal of Higher Education/Revue canadienne d'enseignement sup\'erieur}, 47(3), 101-119.\\
Ashwin, P., \& Case, J. M. (2018). Higher Education Pathways: South African Undergraduate Education and the Public Good (p. 308). African Minds.\\
Bao, L., \& Koenig, K. (2019). Physics education research for $21^{st}$ century learning. Disciplinary and Interdisciplinary Science Education Research, 1(1), 1-12.\\
Bernstein, B. (1999). Vertical and horizontal discourse: An essay. British Journal of Sociology and Education 20: 157-73.\\
Bernstein, B. (2000). Pedagogy, symbolic control, and identity: Theory, research, critique (Vol. 5). Rowman \& Littlefield.\\
Blackie, M. A. 2014. Creating semantic waves: Using Legitimation Code Theory as a tool to aid the teaching of chemistry. Chemistry Education Research and Practice, 15 (4), 462-469. DOI: 10.1039/C4RP00147H\\
Biggs, J., \& Tang, C. (2015). Constructive alignment: An outcomes-based approach to teaching anatomy. In Teaching anatomy (pp. 31-38). Springer, Cham.\\
Boud, D., \& Falchikov, N. (2006). Aligning assessment with long-term learning. Assessment \& evaluation in higher education, 31(4), 399-413.\\
Council on Higher Education (2014). Vital stats: public higher education 2014. Pretoria: Council on Higher Education.\\
Conana, C. H. (2016). Using semantic profiling to characterize pedagogical practices and student learning: A case study in two introductory Physics courses.\\
Conana, H., Marshall, D., \& Case, J. (2020). A SEMANTICS ANALYSIS OF FIRST-YEAR PHYSICS TEACHING. Building Knowledge in Higher Education: Enhancing Teaching and Learning with Legitimation Code Theory.\\
Feynman, R. P. (1965). The Character of Physical Law, BBC.\\
Georgiou, H. (2016). Putting physics knowledge in the hot seat: The semantics of student understandings of thermodynamics. In Maton, K., Hood, S., and Shay, S. (eds.). 2016. Knowledge-building: Educational studies in Legitimation Code Theory. London: Routledge. 176-192.\\
Gibbs, G. (2010). Using assessment to support student learning. Leeds Met Press. ISBN 978-1- 907240-06-5\\
Maton, K. (2014). Knowledge and Knowers: Towards a realist sociology of education, London: Routledge.\\
Maton, K. (2016). Legitimation Code Theory: building knowledge about knowledge-building, in K. Maton, S. Hood \& S. Shay (eds) Knowledge-building: educational studies in Legitimation Code Theory. London, Routledge.\\
Maton, K., \& Doran, Y. J. (2021). Constellating science: How relations among ideas help build knowledge. In K. Maton, J. R. Martin \& Y. J. Doran, Y. J. (Eds.). (2021). Teaching Science: Knowledge, Language, Pedagogy. Routledge.Teaching Science. Abington: Routledge. 49-75.\\
Quan, G. (2017). Becoming a Physicist: How Identities and Practices Shape Physics Trajectories (Doctoral dissertation).\\
Roberts, P. (2015). Higher education curriculum orientations and the implications for institutional curriculum change. Teaching in Higher Education, vol. 20, no. 5, 542-555.\\
Towne, L., \& Shavelson, R. J. (2002). Scientific research in education. National Academy Press Publications Sales Office.\\
Van Heuvelen, A. (1991). Learning to think like a physicist: A review of research-based instructional strategies. American Journal of Physics, 59(10), 891-897.\\
Wang, C. (2018). Scientific Culture and the Construction of a World Leader in Science and Technology. Cultures of Science, 1(1), 1-13.\\
Wisker, G. (2018). Different journeys: Supervisor perspectives on disciplinary conceptual threshold crossings in doctoral learning. Critical Studies in Teaching and Learning, 6(2).\\
Zimmerman, C. (2000). The development of scientific reasoning skills. Developmental review. 20(1), 99-149.

\end{document}